\documentclass[twocolumn,letter]{jpsj3} 
\usepackage{graphicx}
\usepackage{dcolumn}
\usepackage{bm}
\usepackage{color}

\title{Pressure Study of BiS$_2$-Based Superconductors \\
Bi$_4$O$_4$S$_3$ and La(O,F)BiS$_2$}

\author{Hisashi \textsc{Kotegawa}$^{1}$\thanks{E-mail address: kotegawa@crystal.kobe-u.ac.jp}, Yusuke \textsc{Tomita}$^{1}$, Hideki \textsc{Tou}$^{1}$, Hiroki \textsc{Izawa}$^{2}$, Yoshikazu \textsc{Mizuguchi}$^{2}$, Osuke \textsc{Miura}$^{2}$, Satoshi \textsc{Demura}$^{3}$, Keita \textsc{Deguchi}$^{3}$, and Yoshihiko \textsc{Takano}$^{3}$}

\inst{$^{1}$Department of Physics, Kobe University, Kobe 658-8530, Japan\\
$^{2}$Department of Electrical and Electronic Engineering, Tokyo Metropolitan University, 1-1, Minami-osawa, Hachioji, 192-0397, Japan\\
$^{3}$National Institute for Materials Science, 1-2-1, Sengen, Tsukuba 305-0047, Japan\\

}

\abst{
We report the electrical resistivity measurements under pressure for the recently discovered BiS$_2$-based layered superconductors Bi$_4$O$_4$S$_3$ and La(O,F)BiS$_2$. In Bi$_4$O$_4$S$_3$, the transition temperature $T_c$ decreases monotonically without a distinct change in the metallic behavior in the normal state. In La(O,F)BiS$_2$, on the other hand, $T_c$ initially increases with increasing pressure and then decreases above $\sim1$ GPa. The semiconducting behavior in the normal state is suppressed markedly and monotonically, whereas the evolution of $T_c$ is nonlinear.
The strong suppression of the semiconducting behavior without doping in La(O,F)BiS$_2$ suggests that the Fermi surface is located in the vicinity of some instability.
In the present study, we elucidate that the superconductivity in the BiS$_2$ layer favors the Fermi surface at the boundary between the semiconducting and metallic behaviors.
}

\kword{Bi$_4$O$_4$S$_3$, La(O,F)BiS$_2$, pressure}

\begin{document}
\maketitle

The appearance of new superconductors with a BiS$_2$ layer has opened a new route in the field of fascinating layered superconductors.\cite{Mizuguchi_443}
Soon after the first discovery of superconductivity in Bi$_4$O$_4$S$_3$, another system with the same BiS$_2$ layer, La(O,F)BiS$_2$, joined the family of this compound.\cite{Mizuguchi_La}
The discovery of superconductivity in this system is significant in many aspects, because it gives an opportunity for wide and thorough investigations using a variety of related materials possessing the BiS$_2$ layer along with the possible substitution of rare-earth ions as well as Fe-based superconductors.
In fact, isostructural Nd(O,F)BiS$_2$ has also shown superconductivity.\cite{Demura}
This derivative finding revealed that the superconductivity occurring in the BiS$_2$ layer is robust against magnetic elements.
For either Bi$_4$O$_4$S$_3$ and La(O,F)BiS$_2$, superconductivity can be obtained by carrier (electron) doping into the band insulators Bi$_6$O$_8$S$_5$ or LaOBiS$_2$.
The band calculation for the composition Bi$_4$O$_4$S$_3$ suggests that the Fermi level is located at the peak of the density of states, which mainly originates from the Bi $6p$ orbitals.\cite{Mizuguchi_443}
LaOBiS$_2$ also has a similar band structure, and the F-doping of $x=0.5$ corresponds to the Fermi level with a high density of states in the rigid band picture.\cite{Usui}
An important suggestion by the band calculations is that the bands possess a quasi-one-dimensional character, which originates from the difference in the contribution between the $p_x$ and $p_y$ orbitals, for both compounds.
This gives a good nesting property, suggestive of the importance of electron correlations for superconductivity.\cite{Mizuguchi_443,Usui} 
Another notable point is that the Fermi level for the nominal composition is located in the vicinity of the topological change in the Fermi surface.\cite{Usui}
If this superconductivity originates from the electric correlation, this suggests that the superconducting (SC) symmetry can change depending on the doping level.\cite{Usui}

Among the three compounds discovered thus far, the maximum $T_c$ is $\sim10$ K for La(O,F)BiS$_2$.
It is an intriguing issue how a high $T_c$ is possible in the BiS$_2$ layer.
The doping dependence of $T_c$ has been investigated in Nd(O,F)BiS$_2$, and the peak of $T_c$ appears at $x=0.3$ although it is not so sensitive to the doping level.\cite{Demura}
In addition to the change in the doping level, the structural compression by applying pressure is also an effective method of exploring the optimized condition for superconductivity. 
The lattice constants at ambient pressure are different among the compounds: $a=3.9592$ \AA\ for Bi$_4$O$_4$S$_3$ and $a=4.0527$ \AA\ for La(O,F)BiS$_2$.\cite{Mizuguchi_443,Mizuguchi_La} 
$T_c$ is twofold higher in La(O,F)BiS$_2$ than in Bi$_4$O$_4$S$_3$.
This suggests that the lattice parameter is an important factor for superconductivity.

A polycrystalline sample of Bi$_4$O$_4$S$_3$ was prepared by the solid-state reaction method,\cite{Mizuguchi_443} while a polycrystalline sample of LaO$_{0.5}$F$_{0.5}$BiS$_2$ was obtained by a two-step process including high-pressure annealing.\cite{Mizuguchi_La}
The composition expressed here is a nominal one for both compounds.
Electrical resistivity ($\rho$) measurement at high pressures of up to $\sim4$ GPa was carried out using an indenter cell.\cite{indenter}
It was measured by a four-probe method using silver paint for contact.
We used Daphne 7474 as a pressure-transmitting medium.\cite{Murata}
The applied pressure was estimated from the $T_{c}$ of the lead manometer.
The present sample of Bi$_4$O$_4$S$_3$ contains Bi as an impurity, and it shows superconductivity above 2.55 GPa accompanied by a structural phase transition.
We applied pressure of above 2 GPa to Bi$_4$O$_4$S$_3$, but it obstructs the evaluation of $T_c$ intrinsic to Bi$_4$O$_4$S$_3$; thus we omitted the data from this paper.

\begin{figure}[htb]
\centering
\includegraphics[width=0.8\linewidth]{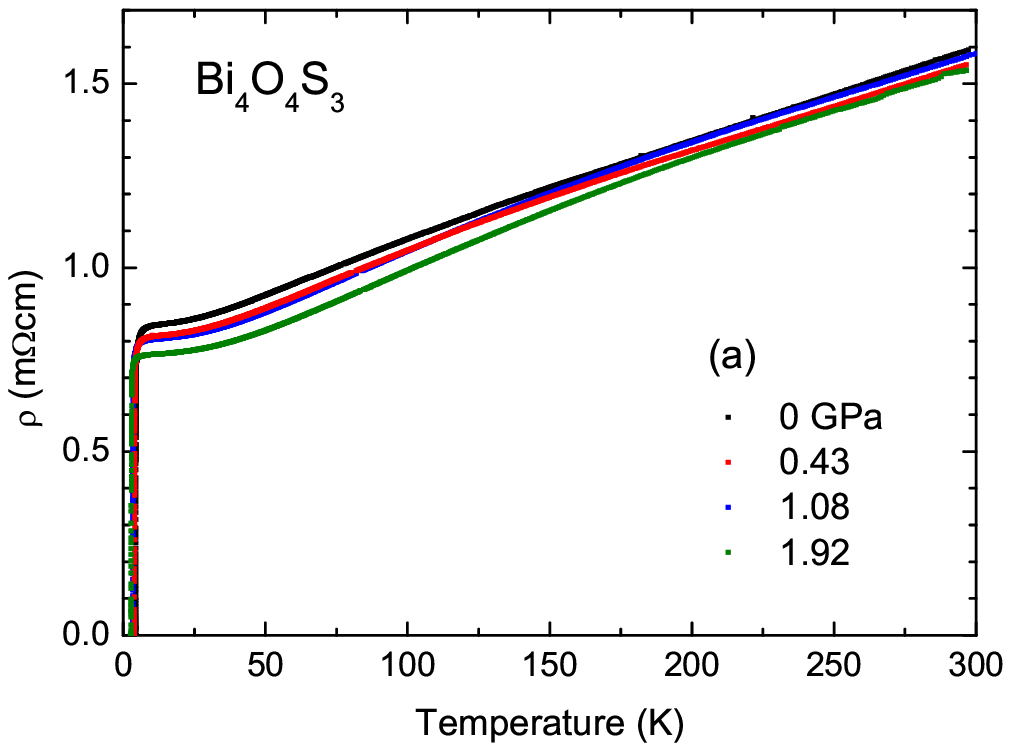}
\includegraphics[width=0.8\linewidth]{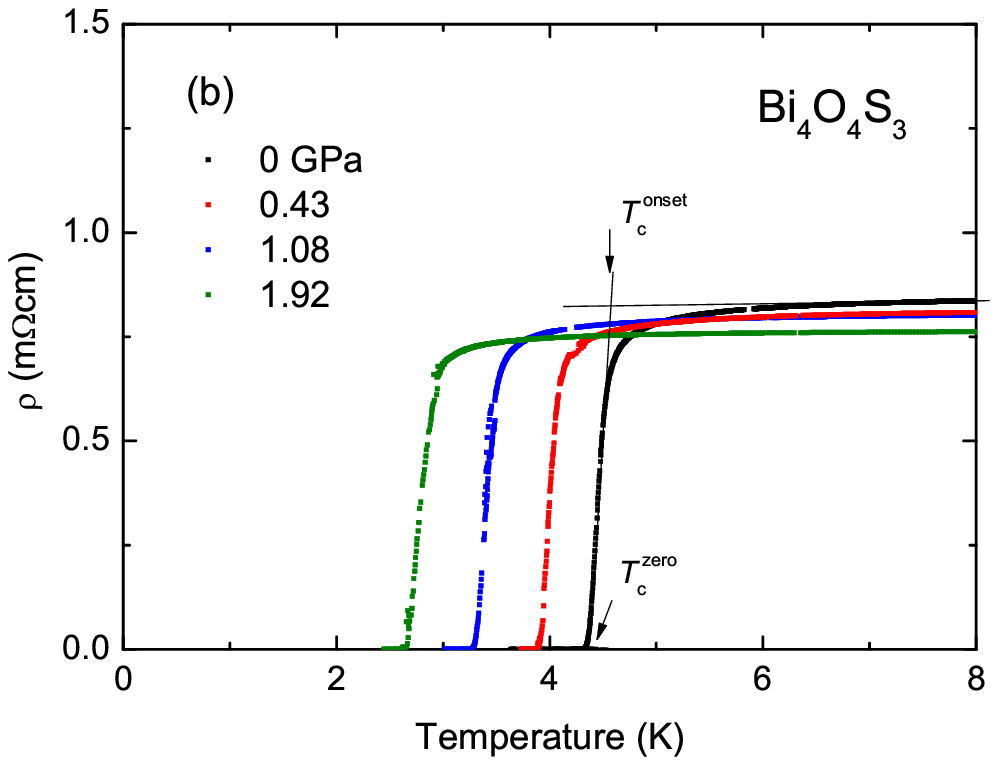}
\caption[]{(color online) (a) Temperature dependence of resistivity for Bi$_4$O$_4$S$_3$ under pressure. It shows the metallic behavior in the normal state, which is not sensitive to pressure. (b) Resistivity at low temperatures. $T_c$ decreases monotonically with increasing pressure. We could not evaluate the intrinsic $T_c$ above 2.5 GPa because of the superconductivity of the Bi impurity. 
}
\end{figure}

\begin{figure}[b]
\centering
\includegraphics[width=0.8\linewidth]{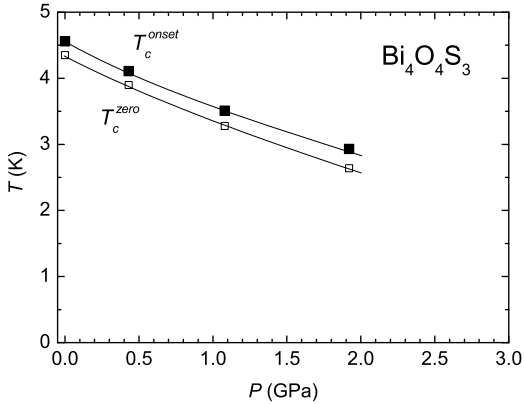}
\caption[]{Pressure dependences of $T_c^{onset}$ and $T_c^{zero}$. Superconductivity for Bi$_4$O$_4$S$_3$ is suppressed under pressure.
}
\end{figure}

Figures 1(a) and (b) show the temperature dependence of resistivity for Bi$_4$O$_4$S$_3$ at pressures of up to 1.92 GPa.
The metallic behavior with a gradual temperature variation is insensitive to pressure.
This behavior in the resistivity might be affected by the contribution of impurity phases; however, similar behavior is observed in the different samples synthesized by other groups.\cite{Li,Tan,Kumar}
On the other hand, the evolution of $T_c$ is noteworthy.
As shown in Fig.~1(b), it decreases sensitively under pressure.
If we determine $T_c^{onset}$ and $T_c^{zero}$ as shown in the figure, $T_c^{onset} \sim 4.6$ K at ambient pressure decreases to 2.9 K at 1.92 GPa.
The pressure dependences of $T_c^{onset}$ and $T_c^{zero}$ are shown in Fig.~2.
It is clear that $T_c$ decreases monotonically.
The initial slope is estimated to be $-1.1$ K/GPa.

\begin{figure}[htb]
\centering
\includegraphics[width=0.8\linewidth]{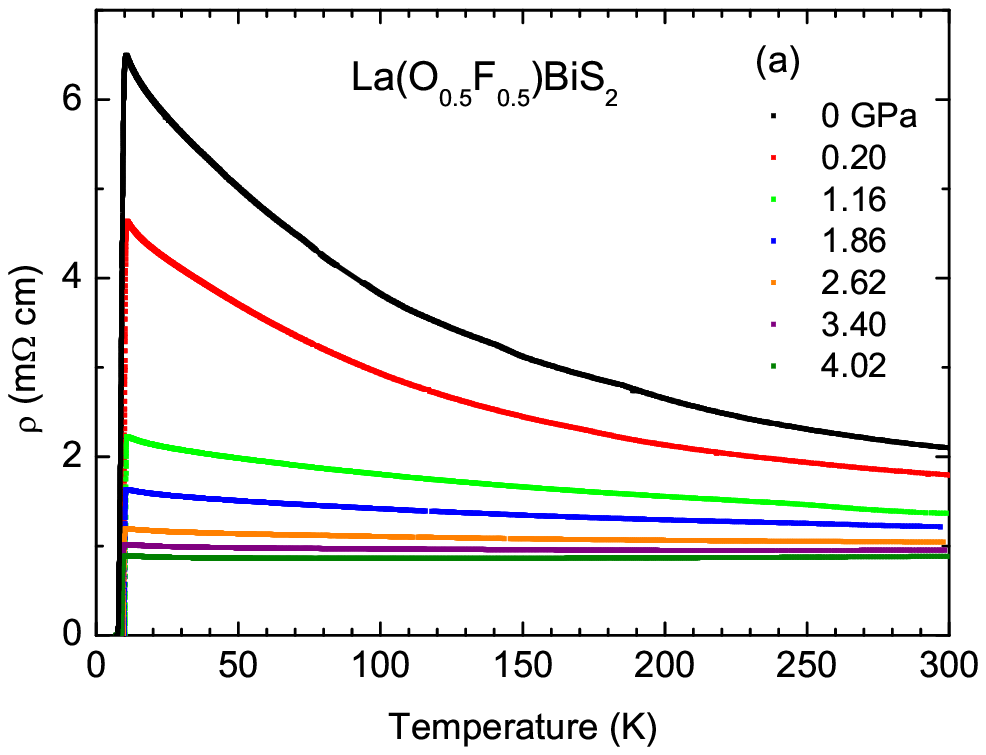}
\includegraphics[width=0.8\linewidth]{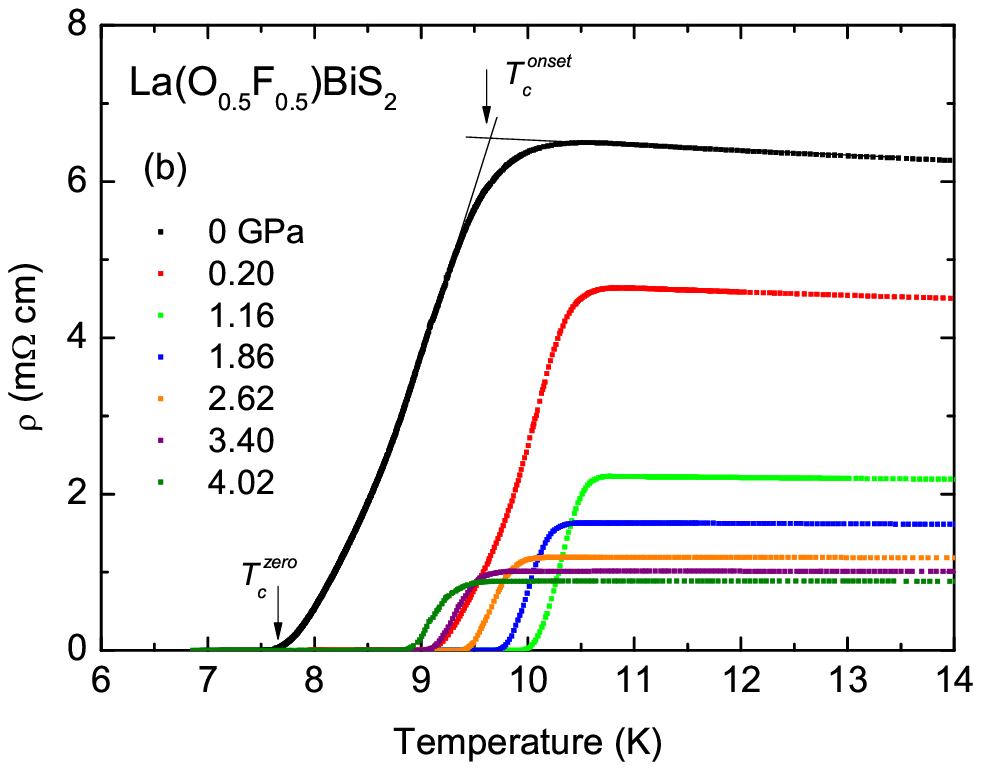}
\caption[]{(color online) (a) Temperature dependence of resistivity for LaO$_{0.5}$F$_{0.5}$BiS$_2$ under pressure. Semiconducting behavior is markedly suppressed with increasing pressure. Almost temperature-independent behavior is observed at $\sim4$ GPa. (b) Resistivity at low temperatures. $T_c$ initially increases and then decreases gradually above 1 GPa.
}
\end{figure}

Figure 3(a) shows the temperature dependence of resistivity for La(O,F)BiS$_2$ below $300$ K measured at several pressures of up to $\sim4$ GPa.
It shows semiconducting behavior at ambient pressure in contrast to that for Bi$_4$O$_4$S$_3$.
This semiconducting behavior is extremely sensitive to pressure and is suppressed under pressure.
We cannot see the obvious semiconducting behavior at $\sim4$ GPa, at which resistivity is almost temperature-independent, which seems to be produced by the comparable contributions of phonon scattering and the semiconducting behavior.
Figure 3(b) shows the data near $T_c$ at low temperatures.
$T_c$ evidently increases by applying pressure up to $\sim1$ GPa, however, it then decreases above $\sim1$ GPa.
Both $T_c^{onset}$ and $T_c^{zero}$, which are determined as shown in the figure, have maxima at approximately 1 GPa; $T_c^{onset}=10.5$ K and $T_c^{zero}=10.0$ K.

\begin{figure}[htb]
\centering
\includegraphics[width=0.65\linewidth]{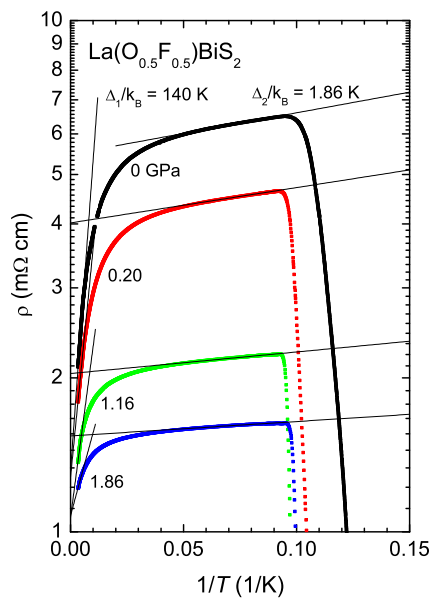}
\caption[]{(color online) $\rho$ vs $1/T$ up to 1.86 GPa. The solid lines give simple activation-type temperature dependences. The respective gaps $\Delta_1$ and $\Delta_2$ are estimated at high and low temperatures.
}
\end{figure}

We attempted to estimate the energy gap assuming that the semiconducting behavior in LaO$_{0.5}$F$_{0.5}$BiS$_2$ originates in the small gap in the band structure.
Figure 4 shows a semi-logarithmic plot for $\rho$ vs $1/T$ up to 2.62 GPa.
The lines in the figure correspond to the simple activation-type relation $\rho \propto \exp(\Delta_{1,2}/k_BT)$.
As seen in the figure, the fitting by a single gap is impossible in the whole temperature range.
$\Delta_1$ is estimated from the fitting in the temperature range of $200-300$ K, while $\Delta_2$ is obtained in the temperature range between $\sim20$ K and $T_c$.
Here, we neglected the contribution of scattering by phonons, but it is comparable to the semiconducting behavior at higher pressures; thus, the estimation of $\Delta_{1,2}$ above 2 GPa was not carried out.
The gap sizes are estimated to be $\Delta_1/k_B\sim140$ K and $\Delta_2/k_B\sim1.86$ K at ambient pressure.
The slopes of both lines in the figure gradually decrease and the semiconducting behavior is suppressed with increasing pressure.

\begin{figure}[htb]
\centering
\includegraphics[width=0.7\linewidth]{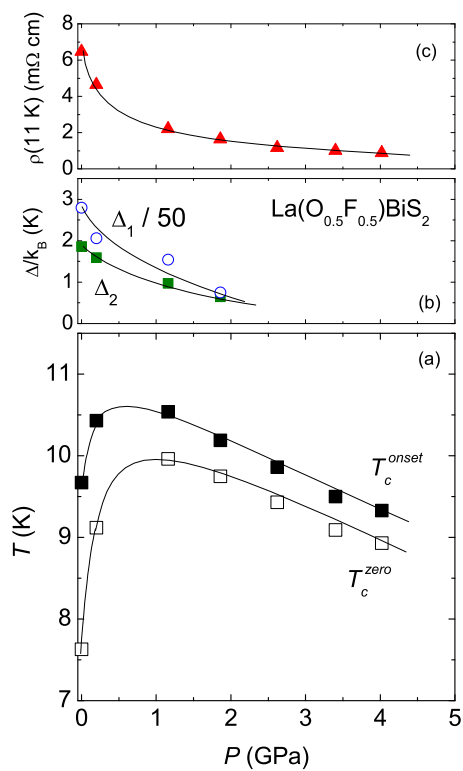}
\caption[]{(color online) (a) Pressure dependences of $T_c^{onset}$ and $T_c^{zero}$ for LaO$_{0.5}$F$_{0.5}$BiS$_2$. $T_c$ initially increases with a relatively large slope and decreases above $\sim1$ GPa. Pressure dependence of (b) tentatively estimated gaps and (c) $\rho$ at 11 K just above $T_c$. The high resistivity induced by the semiconducting behavior is suppressed by applying pressure.
}
\end{figure}

Figure 5 shows the pressure dependences of $T_c^{onset}$ and $T_c^{zero}$ for La(O,F)BiS$_2$, the pressure dependence of $\rho$ at 11 K just above $T_c$, and the tentatively estimated energy gaps $\Delta_{1,2}$.
The high resistivity at low temperatures induced by the semiconducting behavior is markedly suppressed by applying pressure, especially up to $\sim1$ GPa. 
The decrease in $\Delta_{1,2}$ is monotonic.
$T_c$ initially shows a steep increase; however, it gradually decreases above $\sim1$ GPa.
The slope above $\sim 1$ GPa is $\sim -0.4$ GPa/K.
The highest $T_c$ obtained at $0.5-1$ GPa is achieved just after a large suppression of the semiconducting behavior, and $T_c$ monotonically decreases upon approaching the metallic region.

\begin{figure}[htb]
\centering
\includegraphics[width=0.8\linewidth]{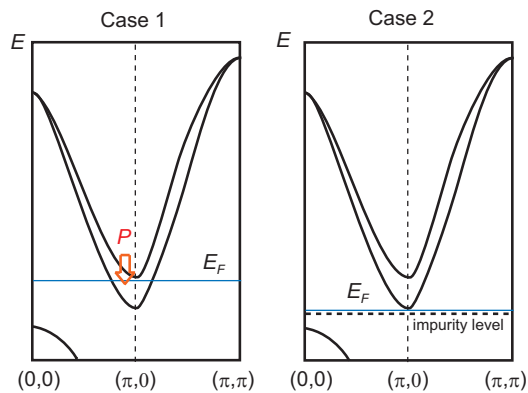}
\caption[]{(color online) Schematic band structures based on calculation for LaOBiS$_2$.\cite{Usui} In Case 1, the Fermi level is located near the position estimated from the nominal composition. The marked change in the resistivity under pressure might be explained by the topological change in the Fermi surface, but it is difficult to explain the semiconducting behavior. In Case 2, conduction is mainly governed by the impurity level or band. Pressure promotes the formation of the impurity band or the overlap of the impurity and conduction bands. 
}
\end{figure}

We first discuss the origin of the strong suppression of the semiconducting behavior in La(O,F)BiS$_2$ under pressure.
When we consider the present pressure effect along with the calculated band structure,\cite{Mizuguchi_443,Usui} we encounter a problem with the actual doping level of electrons.
The band calculation based on the nominal composition has indicated that the Fermi energy is located at the peak position of the density of states in Bi$_4$O$_4$S$_3$.\cite{Mizuguchi_443}
A high density of states at the Fermi level is also expected in LaO$_{0.5}$F$_{0.5}$BiS$_2$ for the nominal composition.\cite{Usui}
Case 1 in Fig.~6 corresponds to the Fermi level close to the estimation from the nominal composition.
If the Fermi level only crosses the lower band near $(\pi,0)$, and the upper band decreases relatively under pressure, then the change in the Fermi surface will give a large evolution of the resistivity.
However this will be in contradiction to the semiconducting behavior in LaO$_{0.5}$F$_{0.5}$BiS$_2$ because the Fermi surface is well established for the doping level of 0.5 electrons/unit cell.
If the semiconducting behavior originates from the thermal change in carrier density, it is naturally conjectured that the carrier density in LaO$_{0.5}$F$_{0.5}$BiS$_2$ is much lower than that expected from the nominal composition, and that the Fermi level is most likely close to the edge of the conduction band.
Here, we conjecture possible origin of the band gap or gap like behavior that explain the semiconducting behavior.
The first one is the original band gap between the conduction band and the valence band.
In the band structure of undoped LaOBiS$_2$,\cite{Usui} the band gap is estimated to be $\sim0.4$ eV, but this tends to be reduced owing to the structural change by F doping, giving a semi metal-like band structure, as shown in Fig.~6.\cite{Kuroki_p}
If the band gap is already small at ambient pressure, thermal excitation from the valence band, which is mainly formed by oxygen orbitals, into the conduction band is possible within the observed temperature range.
Pressure may help to close the band gap.
The second origin is the energy gap between the conduction band and the impurity level originating from the insufficient doping, as shown in Fig.~6 (Case 2).
The third origin is the thermal hopping in the impurity level, which gives gap like behavior.
The semiconducting behavior is robust against magnetic field up to 5 T;\cite{Mizuguchi_La} thus, the tentatively estimated small gap $\Delta_2$ might not originate from a pure band-gap but be attributed in such thermal hopping in Case 2 scenario.
Quantitative discussion is difficult at present; however, the semiconducting behavior in LaOBiS$_2$ seems to suggest the presence of an impurity level.
In such a case, it can be conjectured that applying pressure promotes the formation of the impurity band or the overlap of the impurity and conduction bands, resulting in a decrease in the gaps.

Next, we discuss superconductivity in the BiS$_2$-based systems.
The present pressure experiments suggest that a high $T_c$ is obtained at the boundary between the semiconducting and metallic behaviors: $T_c$ in semiconducting LaO$_{0.5}$F$_{0.5}$BiS$_2$ is twofold higher than that in metallic Bi$_4$O$_4$S$_3$, and once metallic behavior appears, $T_c$ decreases monotonically as seen in LaO$_{0.5}$F$_{0.5}$BiS$_2$ above $\sim1$ GPa and in Bi$_4$O$_4$S$_3$ from ambient pressure.
Similarly, in the isostructural NdO$_{1-x}$F$_{x}$BiS$_2$, $T_c$ is not sensitive to the doping level but its maximum is obtained at $x=0.3$ at which slight semiconducting behavior is observed.\cite{Demura}
These suggest that a high density of states is not important for a high $T_c$ in these systems.
The situation for achieving a high $T_c$ is reminiscent of the doping dependence of Li$_x$ZrNCl, whose electronic specific heat coefficient is $\gamma \sim 1$ mJ/molK$^2$.\cite{Taguchi1}
$T_c$ in Li$_x$ZrNCl increases significantly upon approaching the band insulator.\cite{Taguchi2}
The highest $T_c$ in Li$_x$ZrNCl is achieved when resistivity shows almost temperature-independent behavior.
The evolution of $T_c$ against the doping level cannot be explained by the framework of phonon-mediated superconductivity,\cite{Akashi} and can be interpreted by the nesting scenario.\cite{Kuroki}
If the Fermi surface is well established by sufficient doping in the BiS$_2$-based system, the nesting scenario is a good candidate;\cite{Usui} if not, other mechanisms might be required.

In summary, we have measured the electrical resistivity under pressure for the BiS$_2$ based layered superconductors Bi$_4$O$_4$S$_3$ and LaO$_{0.5}$F$_{0.5}$BiS$_2$.
$T_c$ in metallic Bi$_4$O$_4$S$_3$ decreases monotonically with increasing pressure, whereas $T_c$ in semiconducting LaO$_{0.5}$F$_{0.5}$BiS$_2$ initially increases and then decreases upon approaching the metallic region.
The strong suppression of the semiconducting behavior in LaO$_{0.5}$F$_{0.5}$BiS$_2$ suggests that the Fermi surface is in the vicinity of some instability.
It is plausible that the Fermi level is located near the edge of the conduction band.
This indicates that superconductivity is realized in the complex Fermi surface including the impurity band with a low carrier density and a low density of states, and it rather provides benefit for achieving a higher $T_c$.

\section*{Acknowledgement}

The authors thank Yuki Shimizu for experimental support, and Kazuhiko Kuroki and Hisatomo Harima for helpful discussions.
This work has been partially supported by Grants-in-Aid for Scientific Research (Nos. 22340102, 20102005, and 24340085 from the Ministry of Education, Culture, Sports, Science and Technology (MEXT) of Japan.

\end{document}